\documentclass{article}
\usepackage[utf8]{inputenc}
\usepackage{amsthm}
\usepackage{graphicx,ae,aecompl}
\usepackage{amssymb}
\usepackage{amsmath}
\usepackage{subfigure}
\usepackage{algorithmic}
\usepackage[boxruled,linesnumbered]{algorithm2e}
\usepackage{makeidx}
\usepackage{float}
\usepackage{scalefnt}
\usepackage{indentfirst}

\title{Wavelet Shrinkage in Nonparametric Regression Models with Positive Noise}
\author{Alex Rodrigo dos S. Sousa \\ University of São Paulo, Brazil  \\ \and Nancy L. Garcia \\ State University of Campinas, Brazil \\ }

\begin{document}

\numberwithin{equation}{section}
\numberwithin{table}{section}
\numberwithin{figure}{section}

 \maketitle
    \begin{abstract}
    Wavelet shrinkage estimators are widely applied in several fields of science for denoising data in wavelet domain by reducing the magnitudes of empirical coefficients. In nonparametric regression problem, most of the shrinkage rules are derived from  models composed by an unknown function with additive gaussian noise. Although gaussian noise assumption is reasonable in several real data analysis, mainly for large sample sizes, it is not general. Contaminated data with positive noise can occur in practice and nonparametric regression models with positive noise bring challenges in wavelet shrinkage point of view.  This work develops bayesian shrinkage rules to estimate wavelet coefficients from a nonparametric regression framework with additive and strictly positive noise under exponential and lognormal distributions. Computational aspects are discussed and simulation studies to analyse the performances  of the proposed shrinkage rules and compare them with standard techniques are done. An application in winning times Boston Marathon dataset is also provided.            
    \end{abstract}

\section{Introduction}

\begin{figure}[h]
\centering
\includegraphics[scale=0.60]{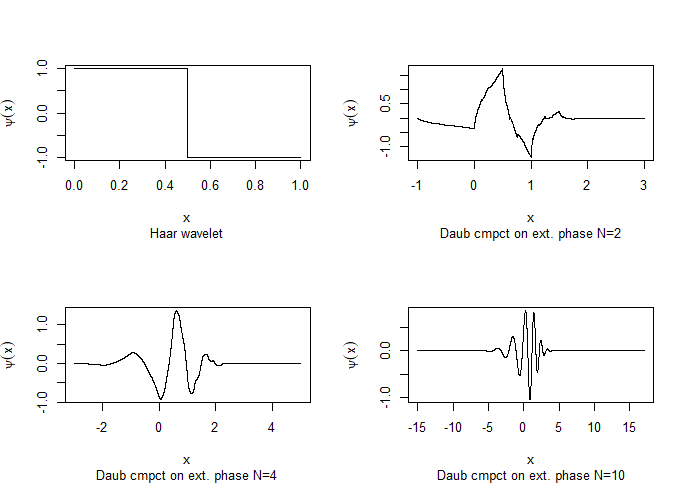}
\caption{Daubechies wavelet functions with $N=1$ (Haar wavelet), $2$ (Daub2), $4$ (Daub4) and $10$ (Daub10) null moments.}\label{fig:wave}
\end{figure}

\indent Wavelet based methods have been applied in several fields of statistics such as time series modelling, functional data analysis, computational methods and nonparametric regression for example. Their success can be justified by several mathematical and computational reasons. In nonparametric regression, the application of this work, it is possible to expand an unknown squared integrable function in orthogonal wavelet basis, which are composed by dilations and translations of a specified function usually called wavelet function or mother wavelet $\psi$. Examples of wavelet functions are Daubechies wavelets, whose are usually indexed by their number of null moments and shown in Figure \ref{fig:wave} for one (Haar or Daub1), two (Daub2), four (Daub4) and ten (Daub10) null moments. This wavelet representation allows the visualization of the data that are obtained from the unknown function by resolution levels and performs a multiresolution analysis by the application of discrete wavelet transform on them. Further, the wavelet representation of a function is typically sparse, i.e, the coefficients of the expansion are majority equal to zero or very close to zero at smooth regions of the represented function domain. This property is important because, once wavelets are well localized in space and frequency domains, the sparsity representation feature provides the identification of the main properties of the unknown function, such as peaks, discontinuities, maximum and minimum by a few amount of nonzero coefficients.  For a review of wavelet methods in statistics, see Vidakovic (1999) and Nason (2008). For a general overview about wavelets and their mathematical properties, see Daubechies (1992) and Mallat (1998).

Wavelet coefficients are essentially sparse at smooth locations of the unknown function, but in practice, after the application of the discrete wavelet transformation on the data, one observes contaminated wavelet coefficients with random noise, called empirical wavelet coefficients, which are not sparse due the noise effect. For denoising the empirical coefficients and estimating the wavelet coefficients of the function representation, thresholding and shrinkage methods are usually applied on the empirical coefficients by reducing their magnitudes. There are several nonlinear thresholding and shrinkage methods available in the literature, most of them based in the seminal works of Donoho (1993a,b), Donoho (1995a,b), Donoho and Johnstone (1994a,b) and Donoho and Johnstone (1995), with the proposition of the so called soft and hard thresholding rules. Bayesian shrinkage procedures have also been successfully proposed for denoising empirical wavelet coefficients. These methods allow the incorporation of prior information regarding to the coefficients, such as the their sparsity, support, dispersion and extreme values by means of a prior probabilistic distribution. In this context, the proposed priors are usually composed by a mixture of a high concentrated distribution around zero to assign sparsity and a symmetric distribution around zero. Prior distributions already proposed to the wavelet coefficients include mixtures of normals by Chipman et al. (1997), mixtures of a point mass function at zero and double exponential distribution by Vidakovic and Ruggeri (2001), Bickel prior by Angelini and Vidakovic (2004), double Weibull by Reményi and Vidakovic (2015), Dirichlet-Laplace priors by Bhattacharya el al. (2015) and, recently, logistic and beta priors by Sousa (2020) and Sousa et al. (2020) respectively. For a general overview abour wavelet shrinkage and thresholding techniques, see Jansen (2001).

Although the well succeeded performance of the proposed thresholding and bayesian shrinkage methods for denoising wavelet coefficients, most of them suppose that the data points from the underlying function are contaminated with additive normal random noise. Despite this assumption can occurs in practice and implies in several good estimation properties, it is not general, mainly under small sample sizes, where central limit theorem can not be applied. Little attention is given for wavelet denoising problems in nonparametric regression models under non-normal random noise or, specifically, additive strictly positive random noise.  Neumann and von Sachs (1995) discuss normal approximations to the wavelet empirical coefficients for thresholding without the normality supposition of the noises and independent and identically distributed (iid) assumption of them. Leporini and Pesquet (2001) proposed the use of Besov priors on the wavelet coefficients to derive a bayesian thresholding rule under a possible resolution level dependent generalized normal distributed noise in the wavelet domain. Antoniadis et al. (2002) provided explicit bayesian thresholding rules based on Maximum a Posteriori (MAP) estimation procedure under exponential power distribution prior on the wavelet coefficients and supposing exponential power and Cauchy distributions to the noise in the wavelet domain. Averkamp and Houdré (2003) analyzed the ideal denoising in the sense of Donoho and Johnstone (1995) considering some classes of noise, including identically distributed symmetric around zero noises in the wavelet domain. Thresholding under compactly support noises in wavelet domain is also discussed. Thus, the above cited works dealt with non-gaussian noise but, no one of them assumes positive noise in the original model. Further, the noise distributions assumtpions occur directly in the wavelet domain, after the discrete wavelet transform application on the original data.   

In this sense, this paper proposes bayesian shrinkage procedures for denoising empirical wavelet coefficients in nonparametric regression models with strictly positive random noise contamination in the original data, assuming additive noises to be independent and identically distributed exponential and lognormal. The adopted priors are the mixture of a point mass function at zero and the logistic prior proposed by Sousa (2020) and beta prior proposed by Sousa et al. (2020), both works under the classical gaussian noise structure. Assuming additive and positive random noise in the original nonparametric model brings several challenges in estimation point of view. First, independent noises property is lost after wavelet transformation, i.e, noises in the wavelet domain are possibly correlated. The consequence of this fact is that the wavelet coefficient estimation can not be performed individually as usually is done under gaussian noise assumption, but jointly by a joint posterior distribution of the wavelet coefficients vector, which requires computational methods, such as Markov Chain Monte Carlo (MCMC) methods to sample from the joint posterior distribution. Further, noises in the wavelet domain are not necessarily positive, but only linear combinations of them. Finally, several statistical models with multiplicative positive noise were proposed and dealt with by logarithmic transformations, but models with additive positive noise are not so common in the literature, although additive positive noise can be observed in a wide variety of real measurements. For example, arrival times of radio or waves measures typically contain positive errors due possibly delays of equipment detection. See Radnosrati et al. (2020) for an interesting study of classical estimation theory of models with additive positive noise and a nice application involving global navigation satellite systems (GNSS) with positive noise arrival times.    

Thus, the main novelty of this work is to perform wavelet shrinkage under additive positive noise in the original nonparametric model. To do so, logistic and beta priors are put on the wavelet coefficients. Logistic prior is suitably for coefficients with support in the Real set. Its scale hyperparameter has easy and direct interpretation in terms of shrinkage, as can be seen in Sousa (2020). The beta prior (Sousa et al., 2020) is a good choice for bounded coefficients and its well known shape flexibility brings advantages in modelling. This paper is organized as follows: the considered statistical models are defined in Section 2 and their associated shrinkage rules with computational aspects described in Section 3. Parameters and hyperparameters choices are discussed in Section 4. Simulation studies to obtain the performance of the shrinkage rules and to compare with standard shrinkage/thresholding techniques are analysed in Section 5. A real data application involving winning times of Boston Marathon is done in Section 6. The paper is concluded with final considerations in Section 7.

\section{Statistical models}
We consider $n=2^J$, $J \in \mathbb{N}$, points $(x_1,y_1),\cdots,(x_n,y_n)$ from the nonparametric regression model
\begin{equation}\label{regmodel}
y_i = f(x_i) + e_i, \hspace{0.5cm} i=1,\cdots,n
\end{equation}
where $f \in \mathrm{L}^2(\mathbb{R}) = \{f:\int f^2 < \infty\}$ is an unknown function and $e_i$ are independent and identically distributed (iid) random noises such that $e_i>0$, $i=1,\cdots,n$. The goal is to estimate $f$ without assumptions about its functional structure, i.e, the estimation procedure will take only the data points into account. In this work, we consider random noise with exponential and lognormal distributions, given by 
\begin{itemize}
\item Exponential distributed noise: $e_i \sim \mathrm{Exp}(\lambda)$
\begin{equation}\label{exp}
h(e_i;\lambda) = \lambda \exp\{-\lambda e_i\}\mathbb{I}_{(0,\infty)}(e_i), \hspace{0.5cm} \lambda > 0,
\end{equation}

\item Lognormal distributed noise: $e_i \sim \mathrm{LN}(0,\sigma)$
\begin{equation}\label{lognor}
h(e_i;\sigma) = \frac{1}{e_i\sigma \sqrt{2\pi}}\exp \Big\{-\frac{\log^2(e_i)}{2\sigma^2}\Big\}\mathbb{I}_{(0,\infty)}(e_i), \hspace{0.5cm} \sigma > 0,
\end{equation}

\end{itemize}

\noindent where $\mathbb{I}_{A}(\cdot)$ is the usual indicator function on the set $A$ and $\log(\cdot)$ is the natural logarithm. We suppose both the noise distribution parameters $\lambda$ and $\sigma$ as known, although a brief discussion for the unknown case is provided in Section 4.

\begin{figure}[h]
\centering
\includegraphics[scale=0.60]{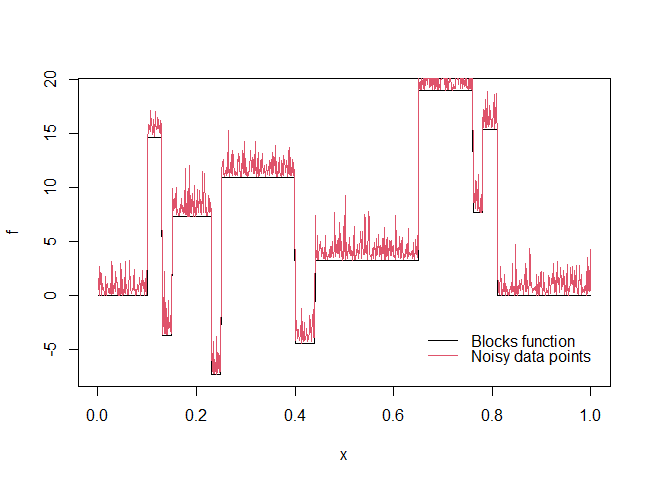}
\caption{Blocks function and 1024 data points with additive exponential noises.}\label{fig:ex}
\end{figure}

The unknown function $f$ can be represented by 
\begin{equation} \label{expan}
f(x) = \sum_{j,k \in \mathbb{Z}}\theta_{j,k} \psi_{j,k}(x),
\end{equation}
where $\{\psi_{j,k}(x) = 2^{j/2} \psi(2^j x - k),j,k \in \mathbb{Z} \}$ is an orthonormal wavelet basis for $L^2(\mathbb{R})$ constructed by dilations $j$ and translations $k$ of a function $\psi$ called wavelet or mother wavelet and $\theta_{j,k}$ are wavelet coefficients that describe features of $f$ at spatial location $2^{-j}k$ and scale $2^j$ or resolution level $j$. In this context, the data points $(x_1,y_1),\cdots,(x_n,y_n)$ can be viewed as an approximation of $f$ at the finest resolution level $J$ with additive and positive noise contamination. As an example, Figure \ref{fig:ex} displays a Donoho-Johnstone (D-J) test function called Blocks, that will be defined in Section 5, and $1024 = 2^{10}$ data points generated from this function with additive exponential distributed random noises.  

The estimation process of $f$ is done by estimation of the wavelet coefficients. In vector notation, model \eqref{regmodel} can be written as
\begin{equation}\label{regvec}
\boldsymbol{y} = \boldsymbol{f} + \boldsymbol{e},
\end{equation}
where $\boldsymbol{y} = [y_1,\cdots,y_n]'$, $\boldsymbol{f} = [f(x_1),\cdots,f(x_n)]'$ and $\boldsymbol{e}= [e_1,\cdots,e_n]'$. A discrete wavelet transform (DWT), which is typically represented by an orthonormal transformation matrix $\boldsymbol{W}_{n \times n} = (w_{ij})_{1 \leq i,j \leq n}$, is applied on both sides of \eqref{regvec}, obtaining the following model in wavelet domain
\begin{equation}\label{wavmodel}
\boldsymbol{d} = \boldsymbol{\theta} + \boldsymbol{\varepsilon},
\end{equation}
where $\boldsymbol{d} = \boldsymbol{Wy}$ is called empirical coefficients vector, $\boldsymbol{\theta} = \boldsymbol{Wf}$ is the wavelet coefficients vector and $\boldsymbol{\varepsilon} = \boldsymbol{We}$ is the random noise vector. Although $\boldsymbol{W}$ is used as DWT representation, fast algorithms are applied to perform DWT in practice, which are more computationally efficient, see Mallat (1998). When $e_i$ is assumed to be iid normal distributed in model \eqref{regmodel}, as occurs in most of the studied nonparametric models in wavelet shrinkage methods research, the distribution of the noise in wavelet domain remains normal, $\varepsilon_i$ is iid normal with the same scale parameter as in the time domain model noise. This property brings several estimation advantages, once the problem of estimating $\theta$ in this context is equivalent of estimating a location parameter of a normal distribution. Moreover, as the noises in wavelet domain remain independent, $\theta$-estimation could be done individually. When $e_i$'s are positive, most of these advantages are lost. Actually, $\varepsilon_i$'s are correlated and not necessarily positive. Also, their distribution is not the same as their counterparts in time domain. The main impact of these facts is that the estimation of $\theta$ can not be performed individually, but according to a joint posterior distribution of $\boldsymbol{\theta}$.     

The wavelet coefficients vector $\boldsymbol{\theta}$ could be estimated by application of a shrinkage rule $\boldsymbol{\delta(d)}$ on the empirical coefficients vector $\boldsymbol{d}$. This procedure essentially performs denoising on the observed coefficients by reducing their magnitudes in order to estimate the wavelet coefficients. After the estimation $\boldsymbol{\hat{\theta}} = \boldsymbol{\delta(d)}$, $f$ is estimated by the inverse discrete wavelet transform (IDWT), $\boldsymbol{\hat{f}} = \boldsymbol{W^{t}\hat{\theta}}$.

\begin{figure}[H]
\subfigure[Logistic densities.\label{lognormal}]{
\includegraphics[scale=0.4]{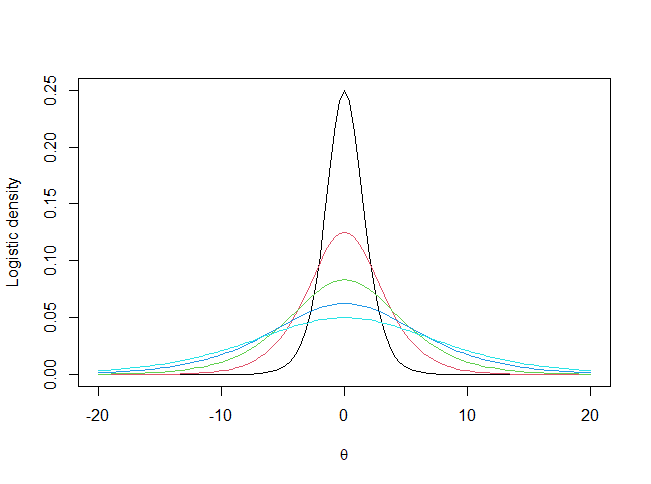}}
\subfigure[Beta densities for $m=3$. \label{blocls}]{
\includegraphics[scale=0.4]{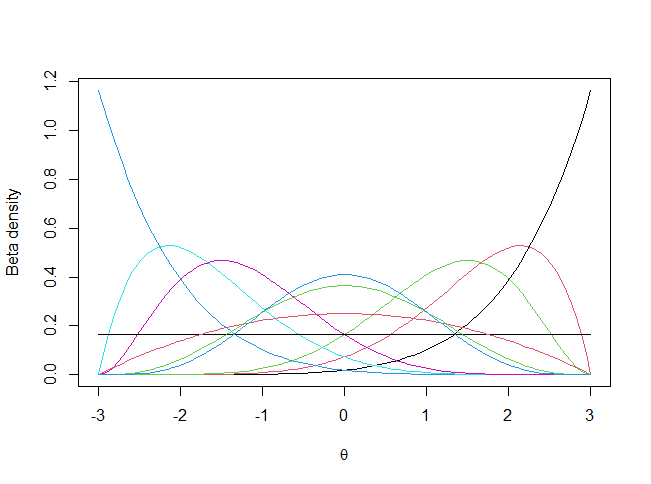}}
\caption{Logistic and beta densities for several hyperparameters values $\tau$ and $(a,b)$ respectively.}\label{fig:densities}
\end{figure}

In this work, we apply a bayesian shrinkage procedure assuming prior distributions to a single wavelet coefficient $\theta$ (the subindices are dropped by simplicity). The priors have the general structure
\begin{equation}\label{prior}
\pi(\theta;\alpha,\boldsymbol{\eta}) = \alpha \delta_{0}(\theta) + (1-\alpha)g(\theta;\boldsymbol{\eta}),
\end{equation} 
for $\alpha \in (0,1)$, $\delta_{0}(\cdot)$ is the point mass function at zero and $g(\cdot;\boldsymbol{\eta})$ is a probability distribution defined according to a hyperparameters vector $\boldsymbol{\eta}$. The choice of $g(\cdot;\boldsymbol{\eta})$ can be made according to the support of $\theta$. We consider in this work two quite flexible distributions $g(\cdot;\boldsymbol{\eta})$, the symmetric around zero logistic distribution proposed by Sousa (2020) given by
\begin{equation}\label{log}
g(\theta;\tau) = \frac{\exp\Big\{-\frac{\theta}{\tau}\Big\}}{\tau \left(1+\exp\Big\{-\frac{\theta}{\tau}\Big\}\right)^2}\mathbb{I}_{\mathbb{R}}(\theta), \hspace{0.5cm} \tau > 0,
\end{equation} 
and the beta distribution on the interval $[-m,m]$ proposed by Sousa et al. (2020) given by
\begin{equation}\label{beta}
g(\theta;a,b,m) = \frac{(\theta+m)^{a-1}(m-\theta)^{b-1}}{(2m)^{a+b-1}B(a,b)}\mathbb{I}_{[-m,m]}(\theta), \hspace{0.5cm} a,b,m>0, 
\end{equation}
where $B(\cdot,\cdot)$ is the beta function. Sousa (2020) and Sousa et al. (2020) developed shrinkage rules under logistic and beta priors respectively under the standard gaussian noise framework. Figures \ref{fig:densities} (a) and (b) show logistic and beta densities for several hyperparameters values respectively. The beta densities are considered on interval $[-3,3]$. 

The logistic prior centered at zero is suitable in bayesian wavelet shrinkage for real valued wavelet coefficients, i.e, when $\theta \in \mathbb{R}$. Further, its hyperparameter $\tau$ has an important role in determining the degree of shrinkage to be applied on the empirical coefficients, as described in Sousa (2020). The beta prior offers great flexibility in modelling bounded wavelet coefficients, i.e, when $\theta \in [-m,m]$, once it allows symmetric ($a=b$) and asymmetric ($a \neq b$) distributions around zero. As the logistic prior, its hyperparameters $a$ and $b$ control the amount of shrinkage of the associated bayesian rule. For $b=a$, bigger values of $a$ imply the increase of the shrinkage level imposed on the shrinkage rule, i.e, the associated rule tends to a severe reduction of the empirical coefficients' magnitudes.  More details about beta priors on wavelet coefficients can be found in Sousa et al. (2020). Thus, logistic and beta priors are convenient choices for $g$ in \eqref{prior} for modelling several prior information about the wavelet coefficients to be estimated, such as their support, symmetry and sparsity.    

\section{Shrinkage rules and computational aspects}
The general shrinkage rules $\delta$ associated to the models \eqref{regmodel}, \eqref{exp}, \eqref{lognor}, \eqref{wavmodel} and \eqref{prior} under quadratic loss function are obtained by the posterior expected value, i.e, $\delta(\boldsymbol{d}) = \mathbb{E}_{\pi}\left(\boldsymbol{\theta}|\boldsymbol{d} \right)$. Once it is infeasible to obtain the posterior expected value analitically, we use an adaptive Markov Chain Monte Carlo (MCMC) method to be described later to generate $L$ samples $\boldsymbol{\theta_1},\boldsymbol{\theta_2},\cdots,\boldsymbol{\theta_L}$ from the joint posterior distribution $\pi(\cdot|\boldsymbol{d})$ of $\boldsymbol{\theta}|\boldsymbol{d}$ and estimate a particular wavelet coefficient $\theta_i$ by the sample mean,
\begin{equation}\label{rule}
\hat{\theta}_i = \delta_i(\boldsymbol{d}) \approx \frac{1}{L}\sum_{l=1}^{L}\theta_{li},
\end{equation}
where $\theta_{li}$ is the $i$-th element of the generated sample $\boldsymbol{\theta_l}$, $l=1,\cdots,L$ and $i=1,\cdots,n$.

The posterior sample generation process is performed using the robust adaptive Metropolis (RAM) algorithm proposed by Vihola (2012) and implemented computationally in the \textit{adaptMCMC} R package by Scheidegger (2021). The algorithm estimates the shape of the target distribution $\pi(\cdot|\boldsymbol{d})$ and simultaneously coerces the mean acceptance rate of process. For each iteration of the chain generation, a single shape matrix $\boldsymbol{S}$ is adaptively updated. Let $\boldsymbol{S_1} \in \mathbb{R}^{n \times n}$ be a lower-diagonal matrix  with positive diagonal elements, $\{\eta_k\}_{k \geq 1} \subset (0,1]$ be a sequence decaying to zero, $\gamma \in (0,1)$ be the target mean acceptance rate and $\boldsymbol{\theta_1}$ such that $\pi(\boldsymbol{\theta_1|\boldsymbol{d}})>0$, the RAM algorithm works as follows for $k \geq 2$,

\begin{enumerate}
\item Generate $\boldsymbol{\theta_{k}^{*}} = \boldsymbol{\theta_{k-1}} + \boldsymbol{S_{k-1}}\boldsymbol{U_k}$,
where $\boldsymbol{U_k} \sim N_n(\boldsymbol{0},\boldsymbol{I_{n}})$ and $\boldsymbol{I_n}$ is the identity matrix of dimension $n \times n$.

\item Do $\boldsymbol{\theta_k} = \boldsymbol{\theta_{k}^{*}}$ with probability 
$$\gamma_k = \min\left(1,\frac{\pi(\boldsymbol{\theta_{k}^{*}}|\boldsymbol{d})}{\pi(\boldsymbol{\theta_{k-1}}|\boldsymbol{d})}\right),$$ 
or $\boldsymbol{\theta_k} = \boldsymbol{\theta_{k-1}}$ else.

\item Computer the lower diagonal matrix $\boldsymbol{S_{k}}$ with positive diagonal elements satisfying the equation
$$ \boldsymbol{S_{k}}\boldsymbol{S_{k}^{t}} = \boldsymbol{S_{k-1}}\left(\boldsymbol{I}+\eta_k(\gamma_k - \gamma)\frac{\boldsymbol{U_k}\boldsymbol{U_{k}^{t}}}{||\boldsymbol{U_k}||^2}\right)\boldsymbol{S_{k-1}^{t}}.$$


\end{enumerate}

We applied $\eta_k = \min\{1,nk^{-2/3}\}$ and $\gamma = 0.234$ as suggested by Vihola (2012) along the simulation studies and application to obtain the posterior distributions samples of the wavelet coefficients. The next subsections provide the posterior distributions that are considered as target distributions in RAM algorithm. 

\subsection{Posterior distributions under exponential noise}
Considering the model under exponential noise \eqref{regmodel}, \eqref{exp} and the model after DWT application   \eqref{wavmodel}, it is straightforward to obtain the likelihood function of the empirical coefficients $\mathcal{L}(\boldsymbol{d} | \boldsymbol{\theta})$ by the application of the Jacobian method to the transformation $\boldsymbol{d} = \boldsymbol{\theta} + \boldsymbol{W}\boldsymbol{e}$. The likelihood function is given by
\begin{equation} \label{likexp}
\mathcal{L}(\boldsymbol{d}|\boldsymbol{\theta}) = |\boldsymbol{W}|\lambda^n\exp\bigg\{-\lambda \sum_i \sum_j w_{ji}(d_j - \theta_j)\bigg\}\prod_i \mathbb{I}_{(0,\infty)}\left(\sum_j w_{ji}(d_j - \theta_j)\right).
\end{equation} 
The posterior distribution of $\boldsymbol{\theta}|\boldsymbol{d}$ can be obtained by the well known relationship 
\begin{equation}\label{bayes}
\pi(\boldsymbol{\theta}|\boldsymbol{d}) \propto \pi(\boldsymbol{\theta}) \mathcal{L}(\boldsymbol{d} | \boldsymbol{\theta}).
\end{equation}
Thus, applying \eqref{bayes} for \eqref{likexp} and the logistic prior model \eqref{prior} and \eqref{log}, we have the following posterior distribution to the wavelet coefficients given the empirical ones under logistic prior model and exponential noise on the original data,
\begin{align}\label{1postlog}
\pi(\boldsymbol{\theta}|\boldsymbol{d}) &\propto \prod_i \left[\alpha \delta_{0}(\theta_i) + (1-\alpha)\frac{\exp\Big\{-\frac{\theta_i}{\tau}\Big\}}{\tau \left(1+\exp\Big\{-\frac{\theta_i}{\tau}\Big\}\right)^2} \right] \times \exp\bigg\{-\lambda \sum_i \sum_j w_{ji}(d_j - \theta_j)\bigg\} \times \nonumber \\ 
& \times \prod_i \mathbb{I}_{(0,\infty)}\left(\sum_j w_{ji}(d_j - \theta_j)\right).
\end{align}

Analogously, we can have the posterior distribution of $\boldsymbol{\theta}|\boldsymbol{d}$ under beta prior model and exponential noise on the original data by considering now \eqref{beta} instead of \eqref{log}, given by
\begin{align}\label{1postbeta}
\pi(\boldsymbol{\theta}|\boldsymbol{d}) &\propto \prod_i \left[\alpha \delta_{0}(\theta_i) + (1-\alpha)\frac{(\theta_i+m)^{a-1}(m-\theta_i)^{b-1}}{(2m)^{a+b-1}B(a,b)} \right] \times \exp\bigg\{-\lambda \sum_i \sum_j w_{ji}(d_j - \theta_j)\bigg\} \times \nonumber \\
& \times \prod_i \mathbb{I}_{[-m,m]}(\theta_i) \times \prod_i \mathbb{I}_{(0,\infty)}\left(\sum_j w_{ji}(d_j - \theta_j)\right).
\end{align}

\subsection{Posterior distributions under lognormal noise}

The likelihood function of the empirical coefficients for the model under lognormal noise \eqref{regmodel}, \eqref{lognor} and the model after DWT application \eqref{wavmodel} is obtained as described in Subsection 3.1 and given by
\begin{align}\label{liklog}
\mathcal{L}(\boldsymbol{d}|\boldsymbol{\theta}) &= \frac{|\boldsymbol{W}|}{(\sigma \sqrt{2 \pi})^{n} \prod_{i} \left[\sum_{j} w_{ji}(d_j - \theta_j) \right]} \times \exp \Bigg \{ -\frac{1}{2\sigma^2} \sum_{i} \log^2\left(\sum_{j} w_{ji}(d_j - \theta_j) \right) \Bigg \} \times \nonumber \\
& \times  \prod_i \mathbb{I}_{(0,\infty)}\left(\sum_j w_{ji}(d_j - \theta_j)\right).
\end{align}

Thus, the posterior distribution of $\boldsymbol{\theta}|\boldsymbol{d}$ under lognormal noise in the original data and logistic prior model \eqref{prior} and \eqref{log} is obtained by application of \eqref{bayes} for the likelihood function \eqref{liklog} and given by 
\begin{align}\label{2postlog}
\pi(\boldsymbol{\theta}|\boldsymbol{d}) &\propto \prod_i \left[\alpha \delta_{0}(\theta_i) + (1-\alpha)\frac{\exp\Big\{-\frac{\theta_i}{\tau}\Big\}}{\tau \left(1+\exp\Big\{-\frac{\theta_i}{\tau}\Big\}\right)^2} \right] \times \frac{\exp \Bigg \{ -\frac{1}{2\sigma^2} \sum_{i} \log^2\left(\sum_{j} w_{ji}(d_j - \theta_j) \right) \Bigg \}}{\prod_{i} \left[\sum_{j} w_{ji}(d_j - \theta_j) \right]} \times \nonumber \\
& \times \prod_i \mathbb{I}_{(0,\infty)}\left(\sum_j w_{ji}(d_j - \theta_j)\right),
\end{align}
and the posterior distribution of $\boldsymbol{\theta}|\boldsymbol{d}$  under beta prior model \eqref{prior} and \eqref{beta} is 
\begin{align}\label{2postbeta}
\pi(\boldsymbol{\theta}|\boldsymbol{d}) &\propto \prod_i \left[\alpha \delta_{0}(\theta_i) + (1-\alpha)\frac{(\theta_i+m)^{a-1}(m-\theta_i)^{b-1}}{(2m)^{a+b-1}B(a,b)} \right]\times \frac{\exp \Bigg \{ -\frac{1}{2\sigma^2} \sum_{i} \log^2\left(\sum_{j} w_{ji}(d_j - \theta_j) \right) \Bigg \}}{\prod_{i} \left[\sum_{j} w_{ji}(d_j - \theta_j) \right]} \times \nonumber \\
& \times \prod_i \mathbb{I}_{[-m,m]}(\theta_i) \times \prod_i \mathbb{I}_{(0,\infty)}\left(\sum_j w_{ji}(d_j - \theta_j)\right).
\end{align}

Therefore, the posterior distributions \eqref{1postlog} and \eqref{1postbeta} of $\boldsymbol{\theta}|\boldsymbol{d}$ are the considered target distributions under logistic and beta prior models respectively in RAM algorithm to be sampled and estimate the wavelet coefficients by the shrinkage rule \eqref{rule} for original data contaminated by exponential noise. Similarly, the posterior distributions \eqref{2postlog} and \eqref{2postbeta} are the target ones under logistic and beta priors respectively for lognormal noise contaminated observations.

\section{Parameters elicitation}

The performance of the bayesian procedure is closely related to a good choice or estimation of the involved parameters and hyperparameters of the models. The proposed shrinkage rules depend on the parameters $\lambda$ and $\sigma$ of the noise exponential and lognormal distributions respectively, which were considered as known throughout the paper, the weight $\alpha$ of the point mass function of the prior models and the hyperparameters $\tau$ and $(a,b,m)$ of the logistic and beta priors respectively.

Angelini and Vidakovic (2004) proposed the hyperparameters $\alpha$ and $m$ be dependent on the resolution level $j$ according to the expressions
\begin{equation}\label{eq:alpha}
\alpha = \alpha(j) = 1 - \frac{1}{(j-J_{0}+1)^r},
\end{equation}
\begin{equation}\label{eq:m}
m = m(j) = \max_{k}\{|d_{jk}|\},
\end{equation}
where $J_ 0 \leq j \leq J-1$, $J_0$ is the primary resolution level, $J$ is the number of resolution levels, $J=\log_{2}(n)$ and $r > 0$. They also suggest that in the absence of additional information, $r = 2$ can be adopted.

The choices of the hyperparameters $\tau$ and $(a,b)$ are discussed respectively by Sousa (2020) and Sousa et al. (2020). In fact, their values have a direct impact on the shrinkage level of the associated rule. Higher denoising level on empirical coefficients requires higuer values of $\tau$ and $(a,b)$. Moreover, these hyperparameters can be resolution level dependent, such as $\alpha$ and $m$. As default values, $\tau = a = b = 5$ can be used. Further discussion about how to choose $(a,b)$ of a beta prior distribution can also be seen in Chaloner and Duncan (1983) and Duran and Booker (1988).

The noise distribution parameters $\lambda$ and $\sigma$ of exponential and lognormal respectively, although considered as known, can be be included in the bayesian framework, independently of the wavelet coefficients, by attributing suitable priors to them, such inverse gamma prior for example. In this case, the general prior model \eqref{prior} under exponential noise could be updated by 
\begin{equation}
\pi(\theta,\lambda;\alpha,\boldsymbol{\eta},\boldsymbol{\zeta}) = \pi(\theta;\alpha,\boldsymbol{\eta}) \times \pi(\lambda;\boldsymbol{\zeta}), \nonumber  
\end{equation}
where $\pi(\lambda;\boldsymbol{\zeta})$ is the prior distribution of $\lambda$ and $\boldsymbol{\zeta}$ is its hyperparameter vector. Analogous procedure can be done for the lognormal noise case.

\section{Simulation studies}
The performances of the proposed shrinkage rules were obtained in simulation studies and compared against standard shrinkage/thresholding tecnhiques. The so called Donoho-Johnstone (D-J) test functions (Donoho and Johnstone, 1995) were considered as underlying functions to be estimated, which are composed by four test functions called Bumps, Blocks, Doppler and Heavisine defined on $[0,1]$ by,
\begin{itemize}
\item \textbf{Bumps}
$$ f(x) = \sum_{l=1}^{11} h_l K\left(\frac{x - x_l}{w_l} \right), \nonumber $$
where 

$K(x) = (1 + |x|)^{-4}$; 

$(x_l)_{l=1}^{11} = (0.1, 0.13, 0.15, 0.23, 0.25, 0.40, 0.44, 0.65, 0.76, 0.78, 0.81)$;

$(h_l)_{l=1}^{11} = (4, 5, 3, 4, 5, 4.2, 2.1, 4.3, 3.1, 5.1, 4.2)$ and 

$(w_l)_{l=1}^{11} = (0.005, 0.005, 0.006, 0.01, 0.01, 0.03, 0.01, 0.01, 0.005, 0.008, 0.005)$.

\item \textbf{Blocks}
$$ f(x) = \sum_{l=1}^{11} h_l K(x - x_l), \nonumber $$
where 

$K(x) = (1 + sgn(x))/2$;

$(x_l)_{l=1}^{11} = (0.1, 0.13, 0.15, 0.23, 0.25, 0.40, 0.44, 0.65, 0.76, 0.78, 0.81)$ and

$(h_l)_{l=1}^{11} = (4, -5, 3, -4, 5, -4.2, 2.1, 4.3, -3.1, 2.1, -4.2)$.

\item \textbf{Doppler}
$$ f(x) = \sqrt{x(1-x)}\sin\left(\frac{2.1 \pi}{x + 0.05} \right). \nonumber$$

\item \textbf{Heavisine}
$$ f(x) = 4\sin(4 \pi x) - sgn(x - 0.3) - sgn(0.72 - x). \nonumber $$

\end{itemize}

\begin{figure}[h]
\centering
\includegraphics[scale=0.60]{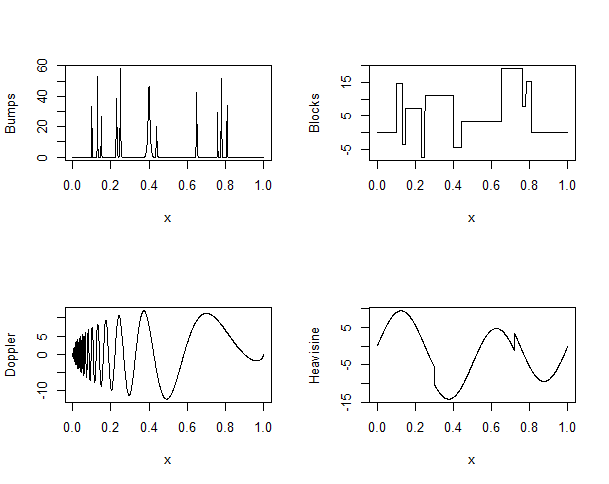}
\caption{Donoho-Johnstone test functions used as underlying signals in the simulation studies.}\label{fig:dj}
\end{figure}

The functions are presented in Figure \ref{fig:dj}. In fact, the D-J functions have important features such as peaks, discontinuities, constant parts and oscillations to be captured by denoising data, representing most of the signals that occur in practice.

For a particular test function, data were generated by adding exponential and lognormal noises to the function points according to two signal to noise ratio (SNR) values, SNR = $3$ and $9$ and two sample sizes, $n = 32$ and $64$. Each scenario of underlying function, SNR and sample size data generation was replicated $M = 100$ times and the averaged mean square error (AMSE) was calculated as performance measure, given by
$$\mathrm{AMSE} = \frac{1}{Mn}\sum_{m=1}^{M} \sum_{i=1}^{n}[{\hat f^{(m)}(x_i)} - f(x_i)]^2, \nonumber$$
where $\hat f^{(m)}(\cdot)$ is the estimate of the function at a particular point in the $m$-th replication, $m = 1, \cdots, M = 100$. For each replication, $L = 10,000$ samples of the posterior distributions \eqref{1postlog}, \eqref{1postbeta}, \eqref{2postlog} and \eqref{2postbeta} were obtained by RAM algorithm and the associated shrinkage rules were calculated by \eqref{rule}. The performances of the shrinkage rules under logistic (LOGISTIC) and beta (BETA) priors were compared against four extensively used shrinkage and thresholding methods, Universal thresholding (UNIV) proposed by Donoho and Johnstone (1994), Cross Validation (CV) proposed by Nason (1996), False Discovery Rate (FDR) proposed by Abramovich and Benjamini (1996) and Stein Unbiased Risk Estimator (SURE) proposed by Donoho and Johnstone (1995) . 

\subsection{Simulation under exponential noise}

Table \ref{tab:amse1} shows the AMSEs of the shrinkage and thresholding rules under exponential noise simulated data. In fact, the proposed shrinkage rules had great performances in terms of AMSE in almost all the scenarios. The shrinkage rule under logistic prior was the best estimator for all the scenarios with sample size $n=32$ and for most of the times when $n=64$, being the best estimator in general. The shrinkage rule under beta prior was the best for Bumps function, SNR=3 and $n=64$ and Blocks, SNR=9 and also $n=64$. Even when beta shrinkage rule was not the best one, its performance was close to the logistic rule in general, being the second best estimator. Moreover, the proposed rules worked much better against the standard rules in some of the cases, for example, for Bumps function, SNR = 9 and $n=32$, the AMSEs of logistic and beta rules were respectively 0.787 and 1.140. The third best estimator in those scenarios was SURE, with AMSE = 6.287, almost 8 times the AMSE of logistic rule. Only for heavisine function and $n=64$ we did not have the proposed rules as the best ones, losing for UNIV and CV methods, but even in these cases, their performances were close to these ones. Finally, it should be noted the good behavior of the rules for low signal to noise ratio, i.e, for SNR=3, which is an evidence of good work for high noise datasets. 

\begin{figure}[H]
\centering
\includegraphics[scale=0.80]{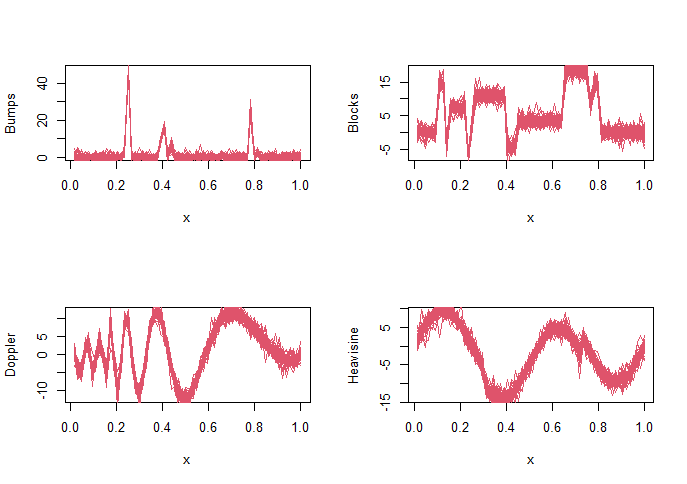}
\caption{Estimates of the D-J test functions by the shrinkage rule under logistic prior in the simulation study for $n=64$, SNR = 9 and for simulated points under exponential noise.}\label{fig:expest}
\end{figure}

\begin{figure}[H]
\centering
\includegraphics[scale=0.80]{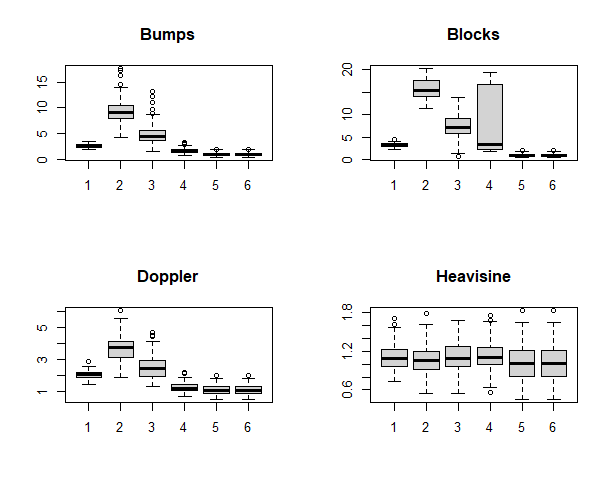}
\caption{Boxplots of the mean square errors (MSE) of the shrinkage and thresholding rules in the simulation study for $n=64$, SNR = 9 and for simulated points under exponential noise. The associated rules are: 1-UNIV, 2-CV, 3-FDR, 4-SURE, 5-LOGISTIC and 6-BETA.}\label{fig:expbp}
\end{figure}

Figure \ref{fig:expest} presents the estimates obtained by the shrinkage rule under logistic prior for $n=64$ and SNR=9. The main features of each test function were captures by the estimates, such as spikes of Bumps, piecewise constant regions of Blocks, oscillations of Doppler and the discontinuity point of Heavisine function. Boxplots of the estimators MSEs are also provided in Figure \ref{fig:expbp} and showed low variation for the proposed shrinkage rules MSEs.  

\begin{table}[H]
\scalefont{0.5}
\centering
\label{my-label}
\begin{tabular}{|c|c|c|c|c|||c|c|c|c|c|}
\hline
Signal & n & Method & SNR = 3 & SNR = 9 & Signal & n & Method & SNR = 3 & SNR = 9   \\ \hline
Bumps& 32	&	UNIV&	18.721	&	2.882	& Blocks	& 32	&	UNIV	&17.631	&	3.292	\\
&	&	CV	&38.439	&	23.175	&	&	&	CV	&21.504	&	15.457	\\
&	&	FDR	&31.603	&	12.530	&	&	&	FDR	&21.684	&	16.227	\\
&	&	SURE &	30.872	&	6.287	&	&	&	SURE	&21.841	&	16.211	\\
&	&	LOGISTIC	& \textbf{7.069}	&	\textbf{0.787}	&	&	&	LOGISTIC &	\textbf{5.960}	&	\textbf{0.748}	\\
& 	&	BETA	&7.081	&	1.140	&	&	&	BETA &	6.542	&	0.769	\\ \hline 
& 64	&	UNIV &	17.052	&	2.615	&	& 64	&	UNIV	&18.002	&	3.211	\\
&	&	CV	&28.317	&	9.140	&	&	&	CV &	24.277	&	16.021	\\
&	&	FDR	&20.496	&	4.562	&	&	&	FDR &	21.586	&	7.864	\\
&	&	SURE &	12.325	&	1.718	&	&	&	SURE &	24.728	&	8.419	\\
&	&	LOGISTIC	&8.449	&	\textbf{1.028}	&	&	&	LOGISTIC &	\textbf{8.303}	&	1.033	\\
&	&	BETA &	\textbf{8.408}	&	1.110	&	&	&	BETA &	8.903	&	\textbf{1.022}	\\ \hline \hline
Doppler&32	&	UNIV	&	11.977	&	1.881	& Heavisine	&32	&	UNIV	&	7.374	&	1.146	\\
&	&	CV	&	12.795	&	3.573	&	&	&	CV	&	7.429	&	1.150	\\
&	&	FDR	&	17.121	&	4.993	&	&	&	FDR	&	7.564	&	1.161	\\
&	&	SURE	&	11.207	&	1.312	&	&	&	SURE	&	7.526	&	1.148	\\
&	&	LOGISTIC	&	\textbf{6.422}	&	\textbf{0.834}	&	&	&	LOGISTIC	&	\textbf{6.373}	&	\textbf{0.779}	\\
&	&	BETA	&	8.488	&	1.109	&	&	&	BETA	&	8.410	&	0.995	\\ \hline 
&64	&	UNIV	&	11.845	&	2.098	&	&	64&	UNIV	&	\textbf{6.425}	&	1.054	\\
&	&	CV	&	12.566	&	3.556	&	&	&	CV	&	6.436	&	\textbf{1.004}	\\
&	&	FDR	&	13.281	&	2.517	&	&	&	FDR	&	6.460	&	1.019	\\
&	&	SURE	&	10.735	&	1.235	&	&	&	SURE	&	6.439	&	1.046	\\
&	&	LOGISTIC	&	\textbf{8.230}	&	\textbf{1.031}	&	&	&	LOGISTIC	&	8.194	&	1.045	\\
&	&	BETA	&	9.780	&	1.124	&	&	&	BETA	&	9.736	&	1.145	\\ \hline

\end{tabular}
\caption{AMSE of the shrinkage/thresholding rules in the simulation study for DJ-test functions under exponential noise.}\label{tab:amse1}
\end{table} 

\subsection{Simulation under lognormal noise}

The obtained results for simulated data under lognormal noise are available in Table \ref{tab:amse2}. In general, the shrinkage rule under logistic prior had the best performance in terms of AMSE, beating the other estimators in practically all scenarios with SNR=9. The rule under beta prior also presented good performance, with AMSEs close to the logistic rule ones and being the best for Blocks function, $n=64$ and SNR=9. Further, the beta rule worked better than logistic one in scenarios with low signal to noise ratio, SNR=3. 

Although logistic rule was the best in general, it should be observed that the behaviors of the standard rules under lognormal noise were better in general than the respective ones under exponential noise. For example, considering data with SNR=3, SURE was the best for Bumps and Doppler underlying functions, while UNIV was the best one for Blocks and Heavisine. Under exponential noise, these rules were dominated by the proposed estimators for these same functions and scenarios.  

\begin{figure}[H]
\centering
\includegraphics[scale=0.80]{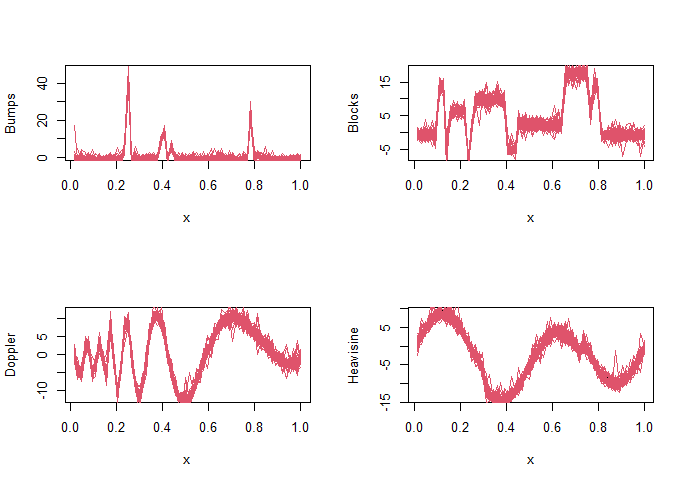}
\caption{Estimates of the D-J test functions by the shrinkage rule under logistic prior in the simulation study for $n=64$, SNR = 9 and for simulated points under lognormal noise.}\label{fig:logest}
\end{figure}

\begin{figure}[H]
\centering
\includegraphics[scale=0.80]{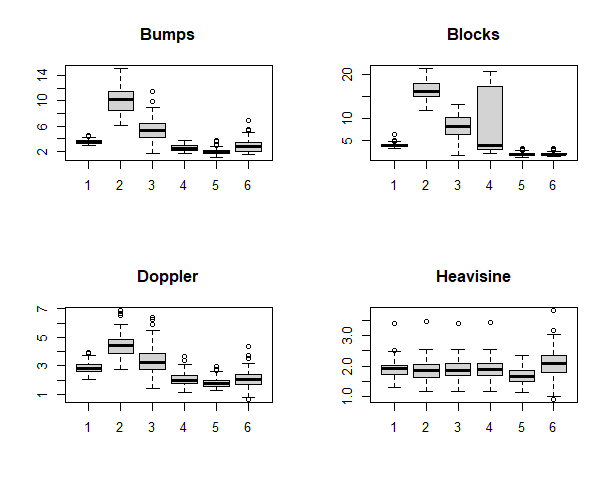}
\caption{Boxplots of the mean square errors (MSE) of the shrinkage and thresholding rules in the simulation study for $n=64$, SNR = 9 and for simulated points under lognormal noise. The associated rules are: 1-UNIV, 2-CV, 3-FDR, 4-SURE, 5-LOGISTIC and 6-BETA.}\label{fig:logbp}
\end{figure}

Figure \ref{fig:logest} shows the estimates of the D-J functions by the shrinkage rule under logistic prior, for $n=64$ and SNR=9. As occured in exponential noise context, the estimates captured well the main characteristics of the test functions. Boxplots of the MSEs are shown in Figure \ref{fig:logbp}, where it is possible to note low MSE variation for the proposed shrinkage rules.

\begin{table}[H]
\scalefont{0.5}
\centering
\label{my-label}
\begin{tabular}{|c|c|c|c|c|||c|c|c|c|c|}
\hline
Signal & n & Method & SNR = 3 & SNR = 9 & Signal & n & Method & SNR = 3 & SNR = 9   \\ \hline
Bumps&32	&	UNIV	&	\textbf{16.940}	&	3.787	& Blocks	&32	&	UNIV	&	\textbf{16.718}	&	4.106	\\
&	&	CV	&	33.612	&	23.744	&	&	&	CV	&	19.909	&	16.154	\\
&	&	FDR	&	25.158	&	13.599	&	&	&	FDR	&	20.177	&	17.023	\\
&	&	SURE	&	24.201	&	6.772	&	&	&	SURE	&	20.306	&	17.049	\\
&	&	LOGISTIC	&	47.405	&	\textbf{2.447}	&	&	&	LOGISTIC	&	45.090	&	\textbf{2.304}	\\
&	&	BETA	&	39.527	&	7.735	&	&	&	BETA	&	50.059	&	2.440	\\ \hline

&64	&	UNIV	&	14.688	&	3.535	&	&	64&	UNIV	&	\textbf{15.721}	&	4.026	\\
&	&	CV	&	20.985	&	9.886	&	&	&	CV	&	20.917	&	16.077	\\
&	&	FDR	&	13.639	&	5.711	&	&	&	FDR	&	15.816	&	8.375	\\
&	&	SURE	&	\textbf{8.441}	&	2.555	&	&	&	SURE	&	20.266	&	7.314	\\
&	&	LOGISTIC	&	27.249	&	\textbf{2.053}	&	&	&	LOGISTIC	&	28.904	&	2.004	\\
&	&	BETA	&	23.539	&	2.958	&	&	&	BETA	&	37.280	&	\textbf{1.970}	\\ \hline \hline

Doppler& 32	&	UNIV	&	9.562	&	2.596	&Heavisine	&32	&	UNIV	&	5.926	&	2.006	\\
&	&	CV	&	9.662	&	3.977	&	&	&	CV	&	\textbf{6.071}	&	2.001	\\
&	&	FDR	&	14.961	&	4.901	&	&	&	FDR	&	6.175	&	2.032	\\
&	&	SURE	&	\textbf{7.603}	& \textbf{1.984}	&	&	&	SURE	&	7.037	&	2.020	\\
&	&	LOGISTIC	&	30.643	&	2.064	&	&	&	LOGISTIC	&	29.421	&	\textbf{1.903}	\\
&	&	BETA	&	15.774	&	3.358	&	&	&	BETA	&	10.235	&	1.926	\\ \hline

&64	&	UNIV	&	9.833	&	2.912	&	&64	&	UNIV	&	\textbf{4.520}	&	1.935	\\
&	&	CV	&	9.749	&	4.517	&	&	&	CV	&	4.647	&	1.895	\\
&	&	FDR	&	9.180	&	3.508	&	&	&	FDR	&	4.888	&	1.926	\\
&	&	SURE	&	\textbf{7.641}	&	2.113	&	&	&	SURE	&	5.463	&	1.940	\\
&	&	LOGISTIC	&	22.200	&	\textbf{1.818}	&	&	&	LOGISTIC	&	20.448	&	\textbf{1.687}	\\
&	&	BETA	&	18.927	&	2.140	&	&	&	BETA	&	14.719	&	2.076	\\ \hline

\end{tabular}
\caption{AMSE of the shrinkage/thresholding rules in the simulation study for DJ-test functions under lognormal noise.}\label{tab:amse2}
\end{table} 

\section{Real data application}

Boston Marathon is one of the most important marathon of the world. It occurs yearly since 1897 with a trajectory of 42,195 Km between Hopkinton and Boston cities, at US Massachussetts state. As mentioned in the introduction, arrival times are classical examples of measurements contaminated by positive noise due possible delays of detection by instruments. 

In this sense, we applied the proposed shrinkage rule with logistic prior under exponential noise assumption for denoising $n=64$ winning times (in minutes) of Boston Marathon Men's Open Division from 1953 to 2016. The data is publicly available at Boston Athletic Association (BAA) webpage \textit{https://www.baa.org/races/boston-marathon/results/champions}. We used a DWT with Daub10 basis and the prior hyperparameters were adopted according to \eqref{eq:alpha} and $\tau = 5$.  

\begin{figure}[H]
\centering
\includegraphics[scale=0.80]{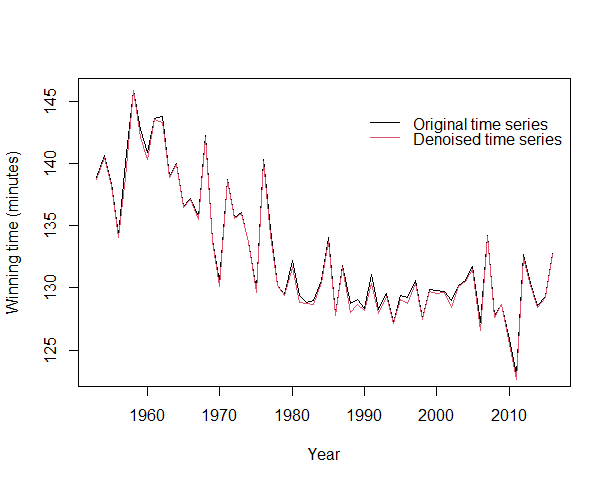}
\caption{Original and denoised winning times of Boston Marathon Men's Open Division between 1953-2016. Denoising was performed by the proposed shrinkage rule with logistic prior under exponential noise model.}\label{fig:app1}
\end{figure}

\begin{figure}[H]
\subfigure[Empirical wavelet coefficients.\label{lognormal}]{
\includegraphics[scale=0.5]{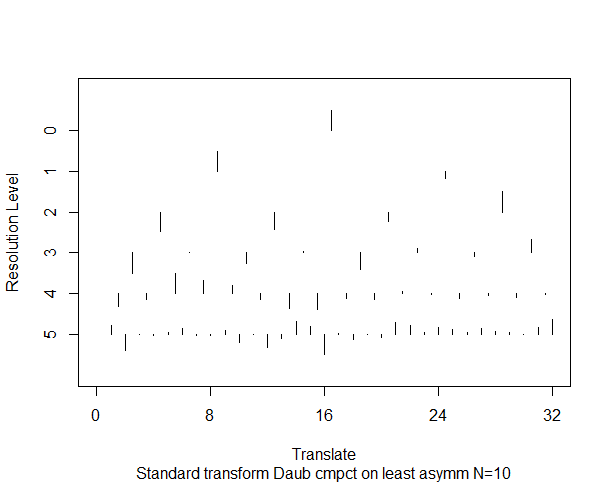}}
\subfigure[Differences between empirical and estimated coefficients. \label{blocls}]{
\includegraphics[scale=0.5]{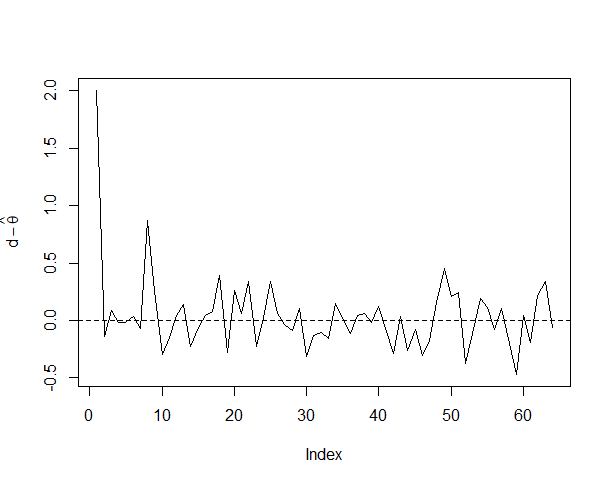}}
\caption{Empirical coefficients by resolution level (a) and differences between empirical and estimated wavelet coefficients (b) of winning times of Boston Marathons dataset. Denoising obtained by application of the shrinkage rule with logistic prior under exponential noise.}\label{fig:app2}
\end{figure}

\begin{figure}[H]
\subfigure[Differences between observed and denoised data (residuals).\label{lognormal}]{
\includegraphics[scale=0.5]{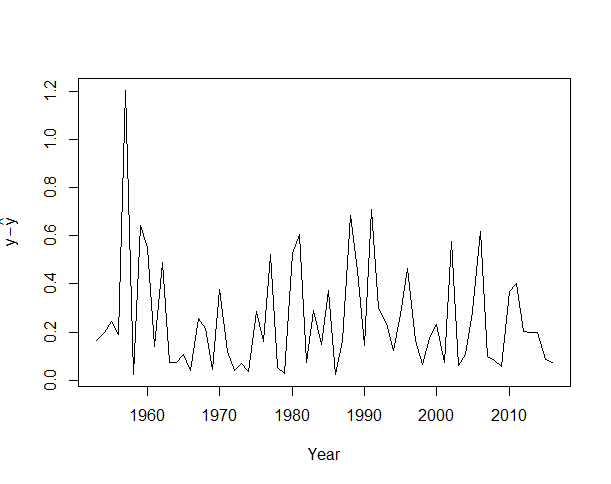}}
\subfigure[Histogram of residuals. \label{blocls}]{
\includegraphics[scale=0.5]{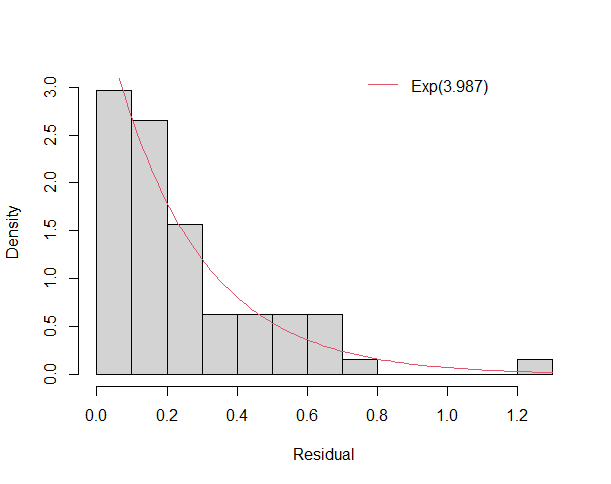}}
\caption{Differences between observed and denoised data (residuals) (a) and histogram of residuals with exponential density curve ($\hat{\lambda} = 3.987$) (b) of winning times of Boston Marathons dataset. Denoising obtained by application of the shrinkage rule with logistic prior under exponential noise.}\label{fig:app3}
\end{figure}

Figure \ref{fig:app1} shows original and denoised data by the shrinkage rule with logistic prior under exponential noise. As expected, the denoised winning times are less than or equal the measured ones, depending on the shrinkage level. Since the good precision of measured times for this competition, it was not necessary the application of a high shrinkage level rule. The empirical wavelet coefficients (represented by vertical bars) by resolution level and the differences between them and the estimated coefficients, $d - \hat{\theta}$, are shown in Figures \ref{fig:app2} (a) and (b) respectively. It is possible to note that, although residuals in original data are positive, which can be seen in Figure \ref{fig:app3} (a), their counterparts in the wavelet domain are not necessarily positive, i.e., there are estimated coefficients bigger than their respective empirical ones.

Finally, Figure \ref{fig:app3} (b) presents the histogram (with area equals to 1) of the residuals in time domain, i.e, $y - \hat{y}$, with a superposed exponential density curve, for $\hat{\lambda} = n/\sum_i (y_i - \hat{y}_i)=3.987$, the maximum likelihood estimate. In fact, the one sample Kolmogorov-Smirnov test for exponential distribution with $\lambda = 3.987$ of the residuals provided a p-value = 0.7057, not rejecting the null hypothesis under $5\%$ of significance level. Thus, the exponential noise assumption for these dataset seems to be reasonable.

\section{Final considerations}
We proposed bayesian wavelet shrinkage rules to estimate wavelet coefficients under nonparametric models with exponential and lognormal additive noise. The adopted priors to the wavelet coefficients were mixtures of a point mass function at zero with logistic and beta distributions. Under the standard gaussian noise assumption, the distribution is preserved on wavelet domain, i.e, the noises after discrete wavelet transform application on original data remain iid gaussian, which allow estimation process coefficient by coefficient. Under positive noise model, this feature is lost. Noises on wavelet domain are not necessarily positive and are correlated. The main impact is that shrinkage is performed on the empirical coefficients vector, which required the application of a robust adaptive MCMC algorithm to calculate posterior expectations, once these are the shrinkage rule under quadratic loss assumption. 

The performances of the proposed shrinkage rules in terms of averaged mean square error (AMSE) were better than standard shrinkage and thresholding techniques in most of the scenarios of the simulation studies. Although the rules are more expensive computationally than the classical methods, their performances in simulation studies can indicate them as promissing shrinkage rules for denoising contaminated data with positive noise. 

The behaviour of the shrinkage rules for other positive support distributed noises and the impact of the wavelet basis choice for performing DWT are suggested as future important questions to be studied in future works.
   
\section*{Acknowledgements}

This work was supported by a CAPES\footnote{Coordination of Superior Level Staff Improvement, Brazil} fellowship. 


\end{document}